\documentclass[aps,prl,twocolumn,showpacs,preprintnumbers,amsmath,amssymb]{revtex4-1}
\pdfoutput=1
\usepackage[breaklinks=true]{hyperref}

\usepackage{color}
\usepackage{graphicx}
\usepackage{dcolumn}
\usepackage{bm}
\usepackage{bbm}

\usepackage{tikz}
\usetikzlibrary{patterns}
\usetikzlibrary{arrows,shapes,positioning}
\usetikzlibrary{decorations.markings}
\usetikzlibrary{decorations.pathmorphing}
\tikzset{snake it/.style={decorate, decoration=snake}}

\def\ep{{\epsilon}}
\def\d{{\rm d}}

\def\frac#1#2{{#1\over #2}}

\def\s{\sqrt}
\def\i{{\rm i}}

\def\CS{{\cal S}}

\def\be{\begin{equation}}
\def\ee{\end{equation}}
\def\ba{\begin{eqnarray}}
\def\ea{\end{eqnarray}}

\def\ddd{\cdot\cdot\cdot}

\def\ep{\epsilon}

\def\BS{\mathbb{S}}
\def\SU{\mathrm{SU}}

\usepackage{braket}

\begin{document}

\title{Holography in de Sitter Space via Chern-Simons Gauge Theory}
\preprint{YITP-21-105; IPMU21-0059}

\author{Yasuaki Hikida,$^a$ Tatsuma Nishioka,$^a$ Tadashi Takayanagi$^{a,b,c}$ and Yusuke Taki$^a$}

\affiliation{$^a$Center for Gravitational Physics, Yukawa Institute for Theoretical Physics, Kyoto University, \\
Kitashirakawa Oiwakecho, Sakyo-ku, Kyoto 606-8502, Japan}

\affiliation{$^b$Inamori Research Institute for Science, 620 Suiginya-cho, Shimogyo-ku, Kyoto 600-8411, Japan}

\affiliation{$^{c}$Kavli Institute for the Physics and Mathematics of the Universe,\\ University of Tokyo, Kashiwa, Chiba 277-8582, Japan}


\begin{abstract}
In this paper we propose a holographic duality for classical gravity on a three-dimensional de Sitter space.
We first show that a pair of SU$(2)$ Chern-Simons gauge theories reproduces the classical partition function of Einstein gravity on a Euclidean de Sitter space, namely $\BS^3$, when we take the limit where the level $k$ approaches $-2$. This implies that the CFT dual of gravity on a de Sitter space at the leading semi-classical order is given by  an SU$(2)$ Wess-Zumino-Witten 
(WZW) model in the large central charge limit $k\to -2$. We give another evidence for this in the light of known holography for coset CFTs. 
We also present a higher spin gravity extension of our duality.
\end{abstract}

\maketitle

\section{Introduction}

Holography has been one of the most promising ideas which provide non-perturbative formulations of quantum gravity \cite{Susskind:1994vu}. 
This approach has been extremely successful for holography in anti-de Sitter space (AdS), namely the AdS/CFT correspondence \cite{Maldacena:1997re}.
However, we are still lacking understandings of holography in de Sitter space (dS), so-called dS/CFT correspondence \cite{Strominger:2001pn,Witten:2001kn,Maldacena:2002vr} (see also \cite{Maldacena:1998ih,Park:1998qk,Park:1998yw}),
though there has been a concrete proposal in four-dimensional higher spin gravity \cite{Anninos:2011ui}, and interesting recent progresses in the light of the dS/dS correspondence \cite{Alishahiha:2004md,Dong:2018cuv,Gorbenko:2018oov,Geng:2021wcq}, holographic entanglement entropy \cite{Ryu:2006bv,Narayan:2015vda,Sato:2015tta,Miyaji:2015yva}, and holography in dS static patch \cite{Susskind:2021dfc,Susskind:2021esx}.
Especially, we are missing the dual conformal field theory (CFT) which lives on the past/future boundary of de Sitter space in Einstein gravity.
In the present article, we hope to present a solution to this fundamental problem for three-dimensional de Sitter space.

The three-dimensional de Sitter space is special in that it is described by a Chern-Simons gauge theory \cite{Witten:1988hc} and that it is expected to be dual to a two-dimensional CFT assuming the standard idea of dS/CFT.
The Chern-Simons description of gravity on $\BS^3$, which is an Euclidean counterpart of de Sitter space, is described by a pair of $\SU(2)$ Chern-Simons gauge theories \cite{Witten:1988hc}. Moreover, it is well-known that an $\SU(2)$ Chern-Simons theory is equivalent to conformal blocks of the $\SU(2)$ Wess-Zumino-Witten (WZW) model \cite{Witten:1988hf}, which has often been regarded as an example of holography. By combining these observations, it is natural to suspect that the gravity on $\BS^3$ and its Lorentzian continuation, i.e., de Sitter space, is dual to $\SU(2)$ WZW model or its related cousins. 

After a little consideration, however, we are immediately led to a puzzle as follows.
Since the classical limit of the Einstein gravity on $\BS^3$ or de Sitter space
is given by the large level limit $k\to \infty$ (see \cite{Banados:1998tb,Castro:2011xb,Castro:2012gc,Cotler:2019nbi,Castro:2020smu,Anninos:2020hfj,Anninos:2021ihe} for various studies of this limit), the central charge $c$ of the dual $\SU(2)$ WZW model at level $k$ approaches to the finite value $c=3k/(k+2)\to 3$  in this limit. On the other hand, the standard idea of dS/CFT \cite{Strominger:2001pn,Maldacena:2002vr} tells us that the classical gravity is dual to the large central charge limit of a CFT.
In what follows, as the main result in this article, we will show that in the large central charge limit $k\to -2$ of the $\SU(2)$ WZW model, the dual Chern-Simons gravity is able to reproduce the results of classical gravity on $\BS^3$.  
Combined this observation with a de Sitter generalization of the conjectured higher spin AdS/CFT duality \cite{Gaberdiel:2010pz}, we will resolve the above puzzle and obtain a concrete dS/CFT in the three-dimensional case.

\section{Chern-Simons Gravity on \texorpdfstring{$\BS^3$}{S3}}
The Einstein gravity on $\BS^3$ is equivalent to two copies of classical $\SU(2)$ Chern-Simons gauge theories, whose action is given by 
\begin{align}\label{CSgaction}
    \begin{aligned}
        I_\text{CSG} &= I_\text{CS} [A] - I_\text{CS} [\bar A]\ , \\
        I_\text{CS}[A]  &=  - \frac{k}{4 \pi} \int_{\mathcal{M}} \text{Tr} \left[ A \wedge d A + \frac{2}{3} A \wedge A \wedge A \right]\ ,  
    \end{aligned}
\end{align}
where $A$ and $\bar{A}$ are the one-form $\SU(2)$ gauge potentials.
The level $k$ is inversely proportional to the three-dimensional Newton constant $G_N$. 
The partition function of a single $\SU(2)$ Chern-Simons theory with a Wilson loop in the spin-$j$ representation (denoted by $R_j$), is given by $\CS^0_j$  \cite{Witten:1988hf}, 
where $\CS$ is the $\CS$-matrix of modular transformation of $\SU(2)$ WZW model:
\ba
    \CS_j^l=\s{\frac{2}{k+2}}\,\sin\left[\frac{\pi}{k+2}\,(2j+1)(2l+1)\right]\ .
\ea
Therefore, the total partition function of the Chern-Simons theory \eqref{CSgaction} for the three-dimensional gravity is evaluated as 
\ba
    Z_\text{CSG}\left[\BS^3,R_j\right]= \big|\CS_0^j\big|^2\ ,  \label{CSrj}
\ea
where we assumed that the Wilson loop is symmetric between the two $\SU(2)$ gauge groups.

Moreover, when two Wilson loops, each in the $R_j$ and $R_l$ representation, are linked, the partition function of the Chern-Simons gravity reads 
\ba
    Z_\text{CSG}\left[\BS^3,L(R_j,R_l)\right] = \big|\CS_j^l\big|^2\ . \label{CSrlj}
\ea
On the other hand, when two Wilson loops are not linked with each other, we obtain 
\ba
    Z_\text{CSG}\left[\BS^3,R_j,R_l\right] = \left|\frac{\CS_0^j\, \CS_0^l}{\CS_0^0}\right|^2\ . \label{CSunl}
\ea

Note that the above partition functions are for the full quantum Chern-Simons theory, and thus we expect they include quantum gravity effects, which will be suppressed in the large $k$ limit.

\section{Holographic Limit for dS/CFT}

Motivated by the standard version of dS/CFT correspondence in \cite{Maldacena:2002vr}, where Einstein gravity limit of three-dimensional de Sitter space is given by the large central charge limit $c\to \i\,\infty$, we argue the following relation between 
the $\SU(2)$ WZW model and the gravity on $\BS^3$:
\begin{align}
    \begin{aligned}
        c &=\frac{3k}{k+2}= \i\,c^{(g)}\ ,\\
        h_j &=\frac{j(j+1)}{k+2}=\i\, h^{(g)}_j\ ,
    \end{aligned}
\end{align}
where $c$ and $h_j$ are, respectively, the central charge and the chiral conformal dimension of a primary field in the $\SU(2)$ WZW model at level $k$, respectively, while the quantities $c^{(g)}$ and $h_j^{(g)}$ are their gravity counterparts and are real valued.
In the gravity, the radius of $\BS^3$, written as $L$, is related to the central charge via the de Sitter counterpart of the well-known relation \cite{Brown:1986nw,Maldacena:2002vr}:
\ba \label{BHf}
    c^{(g)}=\frac{3L}{2G_N}\ .
\ea 
The energy $E_j$ in this gravity dual to the Wilson loop $R_j$ is simply related to the conformal dimension via
\ba
    E_j=\frac{2\,h^{(g)}_j}{L}\ .
\ea

In the semi-classical gravity regime, $L/G_N\gg 1$, we consider
$k\to -2$ limit, which is more precisely described by
\ba
    k=-2+\frac{6\,\i}{c^{(g)}}+O\left(\frac{1}{c^{(g)2}}\right)\ .\label{limkt}
\ea
In addition it is useful to note
\ba
    1-8\,G_N E_j = 1-\frac{24\,h^{(g)}_j}{c^{(g)}}\simeq (2j+1)^2\ .
\ea
Therefore, the Chern-Simons partition function on $\BS^3$ with a single Wilson loop \eqref{CSrj} is evaluated as follows
(in the semi-classical limit $c^{(g)}\gg 1$):
\begin{align}\label{singleg}
    \begin{aligned}
        Z_\text{CSG}\left[\BS^3,R_j\right]
            &\simeq \frac{c^{(g)}}{12}\exp\left[\frac{\pi\, c^{(g)}}{3}\s{1-8 G_N E_j}\right]\ . 
    \end{aligned}
\end{align}
Similarly, the partition function \eqref{CSrlj} on $\BS^3$ with two linked Wilson loops inserted is estimated by 
\begin{align}\label{twowa}
    \begin{aligned}
        &Z_\text{CSG}\left[\BS^3,L(R_j,R_l)\right]  \\
            &\quad\simeq \frac{c^{(g)}}{12} \exp\left[\frac{\pi\, c^{(g)}}{3}\s{1-8G_N E_j}\s{1-8 G_N E_l}\right]\ . 
    \end{aligned}
\end{align}
For unlinked two Wilson lines, we obtain from (\ref{CSunl}):
\begin{align}\label{twoab}
    \begin{aligned}
        &Z_\text{CSG}\left[\BS^3,R_j,R_l\right] \\
            &\quad \simeq \frac{c^{(g)}}{12}\exp\left[\frac{\pi\, c^{(g)}}{3}\left(\s{1-8G_N E_j}+\s{1-8G_N E_l}-1\right)\right]\ .
    \end{aligned}
\end{align}

Notice that in the above we have assumed the limit $k\to -2$, which looks quite different from the semi-classical limit of the Chern-Simons gauge theory.
To see that our new limit gives a correct answer, we will compare the above results with those expected from the direct Einstein gravity calculations in the following.

\section{Gravity on de Sitter Space}

The Euclidean de Sitter black hole solution is given by 
\begin{align}
\d s^2 = L^2 \left[(1\!-\!8G_N E_j\! -\!r^2)\, \d\tau^2\!+\!\frac{\d r^2}{1\!-\!8G_N E_j \!-\!r^2}\!+\!r^2 \d\phi^2\right]\ ,
\end{align}
where $E_j$ is the energy of an excitation \cite{Spradlin:2001pw}.
The black hole horizon is at $r=\s{1-8G_N E_j}$ and the requirement of smoothness 
at the horizon determines the periodicity:
\ba
    \tau\sim \tau+\frac{2\pi}{\s{1-8G_N E_j}}\ .
\ea
On the other hand, the angular coordinate $\phi$ obeys the periodicity $\phi\sim \phi+2\pi$ and there is a conical singularity at $r=0$.
The black hole entropy reads
\begin{align}\label{BHent}
    S_\text{BH} 
    =\frac{\pi\,c^{(g)}}{3}\s{1-8G_N E_j}\ .
\end{align}

It is useful to introduce the coordinate $\theta$ by 
\ba
    r = \s{1-8G_N E_j}\,\sin\theta \qquad \left(0\leq \theta\leq \frac{\pi}{2}\right)\ ,
\ea
which leads to the metric
\begin{align}\label{metsph}
    \d s^2 = L^2 \left[\d\theta^2\!+\!(1\!-\!8G_N\! E_j)(\cos^2\theta\, \d\tau^2\!+\!\sin^2\theta\, \d\phi^2)\right]\ .
\end{align}
Then we evaluate the gravity action 
\ba
    I_\text{G}=-\frac{1}{16\pi G_N}\int \s{g}\,(R-2\Lambda)\ ,
\ea
where $\Lambda=1/L^2$. 
This leads to 
\begin{align}
    I_\text{G} 
        =-\frac{\pi\,c^{(g)}}{3}\s{1-8G_N E_j}\ , 
\end{align}
whose semi-classical gravity partition function $Z_\text{G}=\exp\left[-I_\text{G}\right]$ agrees with the Chern-Simons result 
\eqref{singleg}.

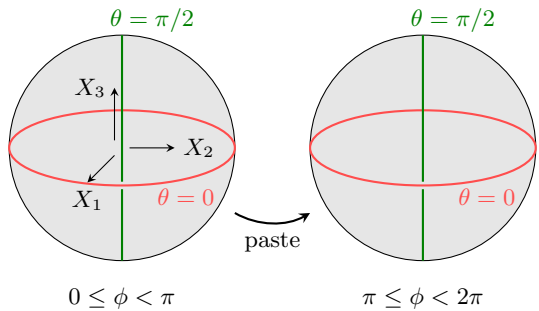
\begin{figure}
  \centering
    \begin{tikzpicture}[>=stealth]
        \draw[fill=gray!20] (-2,0) circle (1.5cm);
        \draw[red!70, thick] (-2, 0) ellipse (1.5cm and 0.5cm);
        \node[red!70] at (-1.15, -0.65) {$\theta=0$};
        \draw[-, green!50!black, thick] (-2, 1.5) -- (-2, -0.45);
        \draw[-, green!50!black, thick] (-2, -0.55) -- (-2, -1.5);
        \node[green!50!black] at (-1.6, 1.7) {$\theta=\pi/2$};
        \draw[->] (-2.1, -0.1) --+ (-0.36, -0.36) node[below] {$X_1$};
        \draw[->] (-1.9, 0) --+ (0.6, 0) node[right] {$X_2$};
        \draw[->] (-2.1, 0.1) --+ (0, 0.7) node[left] {$X_3$};
        \node at (-2, -2) {$0 \leq \phi < \pi$};
        
        \draw[fill=gray!20] (2,0) circle (1.5cm);
        \draw[red!70, thick] (2, 0) ellipse (1.5cm and 0.5cm);
        \node[red!70] at (2.85, -0.65) {$\theta=0$};
        \draw[-, green!50!black, thick] (2, 1.5) -- (2, -0.45);
        \draw[-, green!50!black, thick] (2, -0.55) -- (2, -1.5);
        \node[green!50!black] at (2.4, 1.7) {$\theta=\pi/2$};
        \node at (2, -2) {$\pi \leq \phi < 2\pi$};
        
        \draw[->, thick] (240:1) arc (240:300:1) node[midway, below] {paste};
    \end{tikzpicture}
    \caption{The North [Left] and South [Right] hemisphere with two linked Wilson lines (red and green).}
    \label{wilsc}
\end{figure}

Let us introduce the Cartesian coordinates:
\begin{align}
    \begin{aligned}
        X_1 &= \cos\theta\, \cos\left(\s{1-8 G_N E_j}\,\tau\right)\ ,\\
        X_2 &= \cos\theta\, \sin\left(\s{1-8 G_N E_j}\,\tau\right)\ ,\\
        X_3 &= \sin\theta\, \cos\left(\s{1-8 G_N E_j}\,\phi\right)\ ,\\
        X_4 &= \sin\theta\, \sin\left(\s{1-8 G_N E_j}\,\phi\right)\ .
    \end{aligned}
\end{align}
Then the sphere $\sum_{i=1}^{4}(X_i)^2=L^2$ is described by the metric \eqref{metsph}.
The insertion of the single Wilson line $R_j$ corresponds to a deficit angle $\delta_j=2\pi-2\pi\s{1-8 G_N E_j}$ at $\theta=0$, depicted as the red circle in FIG.\,\ref{wilsc}.

We can realize the second Wilson loop at $\theta=\pi/2$ linking with the first one by identifying the coordinate $\tau$ as
\ba
    \tau\sim \tau+\frac{2\pi\s{1-8G_N E_l}}{\s{1-8 G_N E_j}}\ .
\ea
This is depicted as the green circle in  FIG.\,\ref{wilsc}, where the deficit angle 
$\delta_l=2\pi-2\pi\s{1-8 G_N E_l}$ is present.
Finally, the gravity action for this geometry is estimated as 
\ba
    I_\text{G}=-\frac{\pi\,c^{(g)}}{3}\s{1-8G_N E_j}\s{1-8G_N E_l}\ ,
\ea
which again reproduces the leading part of the Chern-Simons result \eqref{twowa} in the semi-classical limit.

\section{Higher Spin Gravity}

The Chern-Simons theory enables us to construct a broader class of three-dimensional gravity theories, namely, higher spin gravity.
A pair of $\SU(N)$ Chern-Simons theories at level $k$ describes a three-dimensional gravity with spin-$s$ fields for each $s=2,3,\ldots,N$.

The central charge of the $\SU(N)$ WZW model at level $k$ reads
\ba
c=\frac{k(N^2-1)}{k+N}\ .
\ea
The chiral conformal dimension of a primary in the representation specified by a weight vector $\lambda=\sum_{i=1}^{N-1}\lambda_i\, \omega_i $ is given by 
\ba
h_\lambda=\frac{(\lambda,\lambda+2\rho)}{2(k+N)}\ , \label{hsh}
\ea
where $\rho=\sum_{i=1}^{N-1}\omega_i$. The weight lattice is generated by the basis $\{\omega_1,\ddd,\omega_{N-1}\}$ and its inner product is denoted by $(*,*)$. 
Here we follow the convention in \cite{DiFrancesco:1997nk}.
The modular $\CS$-matrix reads
\begin{align}
    \CS_{\lambda}^{\mu} = K\sum_{w\in W}\ep(w)\,\exp\left[-\frac{2\pi\, \i}{k+N}\left(w(\lambda+\rho),\mu+\rho\right)\right]\ ,
\end{align}
where $W$ is the Weyl group and $K$ is a constant fixed by the unitary constraint $\CS\,\CS^\dagger=1$.

Now we analytically continue the level as we did in the $\SU(2)$ case, $c=\i\,c^{(g)}$ and 
$h_\lambda = \i\, h^{(g)}_{\lambda}$, which leads to
\ba \label{hsk1}
    k=-N+N(N^2-1)\, \frac{\i}{c^{(g)}}+O\left(\frac{1}{c^{(g)2}}\right)\ .  \label{limgn}
\ea
Let us evaluate $\CS^0_{0}$, which gives the vacuum partition function $Z_\text{CSG}\left[\BS^3\right]$.
By using the known relation $(\rho,\rho)=N(N^2-1)/12$,
the partition function of the $\SU(N)$ Chern-Simons gravity with linked Wilson loops in the limit \eqref{limgn} looks like:
\begin{align}\label{cft}
    \begin{aligned}
    Z_\text{CSG} &\left[\BS^3 , L (R_\lambda , R_\mu ) \right]
        =  \big|\CS_{ \lambda}^{\mu} \big|^2 \\
        &\quad\sim \exp\left[\frac{\pi\,c^{(g)}}{3}  \frac{(\lambda + \rho , \mu + \rho )}{(\rho , \rho)}\right]\ . 
    \end{aligned}
\end{align}
It is straightforward to confirm that this reproduces the previous result \eqref{twowa} if we set $N=2$.
Moreover, it is useful to note that this group theoretical argument explains the partition function with unlinked Wilson loops
$R_j$ and $R_l$, given by \eqref{CSunl}. Indeed, by setting $\lambda=\lambda_j+\lambda_l$ and $\mu=0$, we can rewrite 
$(\lambda+\rho,\mu+\rho)=(\lambda_j+\rho,\rho)+(\lambda_l+\rho,\rho)-(\rho,\rho)$.

As in the $N=2$ case, we will show below that the 
partition function \eqref{cft} of the $\SU(N)$ Chern-Simons gravity computed from the $k\to -N$ limit of the $\SU(N)$ WZW model equals that of the corresponding higher spin gravity in the classical limit, i.e., the large level limit.
The configuration of the $\SU(N)$ gauge fields describing a conical geometry can be constructed in a similar manner to the AdS case presented in \cite{Castro:2011iw}.
We find it convenient to use the $\bar A = 0$ gauge, where the solution of $A$ is given by
\begin{align}
    A = (h\,b^2\,\bar h)^{-1}\,\d (h\,b^2\,\bar h) \ ,
\end{align}
with parameters:
\begin{align}
    \begin{aligned}
        b &= 
            \prod_{i=1}^N\,\exp\left[ \rho_i\,e_{i,i}\right] \qquad \left(\rho_i \equiv \frac{N+1}{2} - i\right)\ ,\\
        h &=  
            \prod_{i=1}^{\lfloor\frac{N}{2}\rfloor}\exp\left[ - (e_{2i-1,2i} - e_{2i-1, 2i})\,(n_i\,\phi + \tilde n_i\,\tau)\right] \ ,\\
        \bar h &=   
            \prod_{i=1}^{\lfloor\frac{N}{2}\rfloor}\exp\left[ (e_{2i-1,2i} - e_{2i-1, 2i})\,(n_i\,\phi - \tilde n_i\,\tau)\right] \ .
    \end{aligned}
\end{align}
Here $e_{i,j}$ are $N \times N$ matrices with elements $(e_{i,j})_k^{~l} = \delta_{ik}\, \delta_j^{~l}$.

The on-shell action \eqref{CSgaction} for the gauge configuration can be evaluated as
\begin{align}\label{conicalaction}
    I_\text{CSG} = - \frac{\pi}{G_N}\, \frac{\sum_{i=1}^{\lfloor \frac{N}{2}\rfloor}n_i \tilde n_i}{(\rho, \rho)} \ ,
\end{align}
where we use the relation between the Chern-Simons level and the Newton constant in the higher spin gravity:
\begin{align} \label{hsk2}
	k = \frac{L}{8\, G_N\, (\rho, \rho)}\ .
\end{align}
Let us rewrite the eigenvalues as $n_1 \geq n_2 \ldots $, $\tilde n_1 \geq \tilde n_2 \ldots $  and set $n_i = - n_{N+1 - i} \, , \quad 	\tilde n_i = - \tilde n_{N+1 - i}\ $ for $i > \lfloor\frac{N}{2}\rfloor $ \footnote{As in the case of Euclidean AdS$_3$ analyzed in \cite{Castro:2011iw}, we could relax the condition for Lorentzian dS$_3$.}. If we require $n_i \neq n_j$  and $\tilde n_i \neq \tilde n_j$  for $i \neq j$, which generically corresponds to the diagonalizability of the matrix, then 
we could set $n_i =  \lambda_i  + \rho_i \ ,\   \tilde 	n_i = \mu_i  +  \rho_i $  with $\lambda_i  , \mu_i = 0,1,2,\ldots$.
In this representation, with the identification \eqref{BHf}, we can rewrite \eqref{conicalaction} as
\begin{align}
	I_\text{CSG} = - \frac{\pi\, c^{(g)} }{3} \frac{( \lambda + \rho ,  \mu + \rho)}{(\rho,\rho)}\  . \label{bulk}
\end{align}
Hence the on-shell partition function $Z_\text{CSG} = e^{-I_\text{CSG}}$ for the higher spin gravity agrees with the expression \eqref{cft} obtained from the modular $\CS$-matrix as we promised.

\section{Entanglement/Black hole Entropy}
Let us turn to the calculation of entanglement entropy in the gravity on $\BS^3$.
We choose a subsystem $A$ to be a disk on the surface $\BS^2$, which separates $\BS^3$ into two hemispheres. 
We write the boundary circle of $A$ as $\Gamma_A$.
In the replica calculation of entanglement entropy, we introduce a cut along $\Gamma_A$ on $\BS^3$ and take its $n$-fold cover to obtain $\mbox{Tr}[(\rho_A)^n]$.
The replica calculation in Chern-Simons theory was performed in \cite{Dong:2008ft} to read off the topological entanglement entropy \cite{Kitaev:2005dm,Levin:2006zz} in terms of modular matrices. 
In the presence of a Wilson loop $R_\mu$, which is linked with $\Gamma_A$, we obtain (refer to \cite{McGough:2013gka} for an AdS counterpart)
\ba
S_A=\log \big|\CS_{0}^{\mu}\big|^2=\frac{\pi\,c^{(g)}}{3}\frac{(\rho,\mu+\rho)}{(\rho,\rho)}\ .
\ea
For the Einstein gravity ($N=2$) with a Wilson loop $R_j$, it takes the following form: 
\ba
S_A=\log \big|\CS_{0}^{j}\big|^2=\frac{\pi\,c^{(g)}}{3}\s{1-8G_NE_j}\ .
\ea
This indeed coincides with the de Sitter black hole entropy \eqref{BHent}.
It is straightforward to extend the above result to the topological pseudo entropy \cite{Nakata:2020luh,Nishioka:2021cxe}.

\section{Discussions: dS/CFT Interpretation}

We have shown that the limit $k\to -2$ for two copies of the $\SU(2)$ Chern-Simons gauge theories, where the central charge of its dual $\SU(2)$ WZW model gets infinitely large $c^{(g)}\to \i\,\infty$, reproduces the Einstein gravity on $\BS^3$.
More generally, the large central charge limit $k\to -N$ of the $\SU(N)$ WZW model corresponds to the classical limit of the $\SU(N)$ higher spin gravity on $\BS^3$.
We argue that this is a manifestation of the (Euclidean version of) dS/CFT correspondence.

One may worry that this might contradict with the standard fact that the classical limit of  higher spin gravity on $\BS^3$ is given by not finite $k$, but the large $k$ limit of two copies of $\SU(N)$ Chern-Simons theory.
To reconcile this tension, let us consider the following coset CFT, called the $W_N$-minimal model \footnote{The coset realization of $W_N$-minimal model was proven in \cite{Arakawa:2018iyk} and the coset \eqref{cosetc} with generic $k$ was shown to be equivalent to Toda field theory with $W_N$-symmetry in \cite{Creutzig:2021ykz}.}:
\ba
    \frac{\SU(N)_k\times \SU(N)_1}{\SU(N)_{k+1}}\ ,  \label{cosetc}
\ea
which has the central charge 
\ba
c=(N-1)\left(1-\frac{N(N+1)}{(N+k)(N+k+1)}\right)\ .  \label{centwm}
\ea
In \cite{Gaberdiel:2010pz}, this model is argued to be dual to the higher spin gravity on AdS$_3$ (Vasiliev theory \cite{Prokushkin:1998bq}) with two complex scalar fields if we take the 't Hooft limit:
\ba
N\to\infty\ ,\quad k\to \infty\ , \quad \hat{\lambda}=\frac{N}{N+k}=\mbox{fixed}\ .
\ea
This higher spin gravity has the symmetry $\text{hs}[\hat{\lambda}]$, which enhances to $W_\infty[\hat{\lambda}]$ at the asymptotic boundary \cite{Henneaux:2010xg,Campoleoni:2010zq}.
In our limit $k\to -N$, the 
contribution to the total central charge of the coset is dominated by 
the $\SU(N)_k$ part and thus the leading contribution comes from the 
(non-chiral) $\SU(N)$ WZW model, which is essentially the same model we have studied in the above.
Interestingly, the triality \cite{Gaberdiel:2012ku} relates three different values of the previous parameters $(k,N,\hat{\lambda})$ via the following two duality relations:
\begin{align}
    \begin{aligned}
        &\text{(i)} & (k',N',\hat{\lambda}') &= \left(-2N-k-1,\, N,\, -\frac{N}{N+k+1}\right)\ ,\\
        &\text{(ii)} & (k',N',\hat{\lambda}') &=\left(\frac{1-N}{N+k},\, \frac{N}{N+k},\, N\right)\ .
    \end{aligned}
\end{align}
If we apply the duality (ii) to the $k\to -2$ limit (\ref{limkt}) at $N=2$ (see also \cite{Ouyang:2011fs,Perlmutter:2012ds} for a similar continuation),
we find 
\ba
    && (k',N',\hat{\lambda}')\simeq \left(\i\,\frac{c^{(g)}}{6},\, -\i\,\frac{c^{(g)}}{3},\, 2\right)\ .
\ea
Thus this theory has $W_\infty[2]$ symmetry, i.e., Virasoro symmetry, which is indeed expected for the Einstein gravity.
We can generalize this to the $k\to -N$ limit of the $\SU(N)$ theory, for which the duality (ii) predicts $W_\infty[N] \simeq W_N$ symmetry with the level infinitely large as expected for the classical $\SU(N)$ higher spin gravity.
In this way, our dS/CFT example is consistent with an extension of earlier results, at least in the leading order.
It will be interesting future problems to examine correlation functions, quantum gravity corrections, and a Lorentzian continuation explicitly.
We plan to come back to these problems soon \cite{long}.

\vspace{5mm}
{\bf Acknowledgements} 
We are grateful to Yasunori Nomura
for useful discussions.
This work was supported by JSPS Grant-in-Aid for Scientific Research (A) No.\,21H04469, 
Grant-in-Aid for Transformative Research Areas (A) No.\,21H05182, No.\,21H05187 and No.\,21H05190.
T.\,T. is supported by the Simons Foundation through the ``It from Qubit'' collaboration, Inamori Research Institute for Science and 
World Premier International Research Center Initiative (WPI Initiative) from the Japan Ministry of Education, Culture, Sports, Science and Technology (MEXT).  The work of Y.\,H. was supported in part by the JSPS Grant-in-Aid for Scientific Research (B) No.19H01896. The work of T.\,N. was supported in part by the JSPS Grant-in-Aid for Scientific Research (C) No.19K03863.

\bibliography{CSdS}

\end{document}